\documentclass[aps,pra,twocolumn,groupedaddress,showpacs]{revtex4-2}

\usepackage{graphicx}
\usepackage{dcolumn}
\usepackage{bm}
\usepackage{amsmath}
\usepackage{amssymb}
\usepackage{xcolor}
\usepackage{fontenc}
\usepackage{mathrsfs}
\usepackage{float}
\usepackage{natbib}

\def\bra#1{{\left\langle#1\right\vert}}

\def\ket#1{{\left\vert#1\right\rangle}}
\def\braket#1#2{{\left\langle#1\vert#2\right\rangle}}
\def\abs#1{{\left|#1\right|}}

\newcommand{\vac}{\ket{\text{vac}}}

\newcommand{\stkout}[1]{\ifmmode\text{\sout{\ensuremath{#1}}}\else\sout{#1}\fi}

\begin{document}

\title{Photon--Matter Quantum Correlations in Spontaneous Raman Scattering} 

\author{Kai Shinbrough$^1$, Yanting Teng$^{1}$, Bin Fang$^{1}$, Virginia O. Lorenz$^1$, Offir Cohen$^{1,2}$}

\date{\today}

\address{$^1$Department of Physics, University of Illinois at Urbana-Champaign, 1110 West Green Street, Urbana, IL 61801, USA}
\address{$^2$Materials Research Laboratory, University of Illinois at Urbana-Champaign, 104 South Goodwin Avenue, Urbana, IL 61801, USA}

\date{\today}

\begin{abstract}
We develop a 
Hamiltonian formalism 
to study energy and position/momentum correlations between a single Stokes photon and a single material excitation that are created as a pair in the spontaneous Raman scattering process. Our approach allows for intuitive separation of the effects of spectral linewidth, chromatic dispersion, and collection angle on these correlations, and we compare the predictions of the model to experiment. These results have important implications for the use of Raman scattering in quantum protocols that rely on spectrally unentangled photons and collective excitations.

\end{abstract}

\maketitle

\section{Introduction}
Raman scattering is one of the most fundamental light-matter interactions: an incident photon scatters inelastically in a medium, transferring energy to, or gaining energy from, a specific excited state. The lifetime of this excited state, which is finite due to interaction with the environment, dictates the Raman gain spectrum and affects the spatio-temporal structure of the Raman-generated optical field~\cite{Raymer_1985,raymer_stimulated_1981,Raymer1989,Wasilewski_2006} as well as its intensity and fluctuations~\cite{Raymer:82,Walmsley_1983,Raymer_1985,Bogolubov_1987,Fabricius_1984}. The equations of motion for the optical field and medium excitation generated in the Raman interaction have traditionally been solved in the Heisenberg picture, where the temporal decay of the material excitation is taken into account through a dissipation-fluctuation mechanism~\cite{Raymer_1985,raymer_stimulated_1981,Fabricius_1984,scully_zubairy_1997,raymer_stimulated_1981}. Using this formalism an extensive body of work has formed around exploration of the quantum properties of the spontaneously-initiated optical field, 
including the decomposition of the field into independent temporal coherence modes~\cite{Raymer_1985,Raymer1989} and decomposition of the excitation field into corresponding orthonormal spatial modes~\cite{Wasilewski_2006,Raymer04}. Here we investigate the quantum correlations (entanglement) between modes of a single Stokes photon and its single material excitation counterpart in the  Schr\"{o}dinger picture, focusing instead on the spectral representation of these modes. Motivating this work is the necessity of pure, uncorrelated photonic quantum states for many quantum applications, which, in the case of Raman scattering~\cite{DLCZ,Kimble,GisinCrypto}, occurs when no correlations exist between the Stokes photon and medium excitation, apart from their coexistence.

We begin by incorporating the environment degrees of freedom into the system Hamiltonian, and writing the Raman interaction in terms of the eigenmodes of the medium, which includes the environment. 
Our approach then provides an intuitive understanding of photon--matter correlations arising due to energy and momentum conservation in the same way those correlations arise in the photon--photon pairs created in spontaneous parametric down-conversion (SPDC)~\cite{Grice97} or spontaneous four-wave mixing (SFWM)~\cite{garay-palmett_2007}. Our state formalism reveals photon--matter correlations that have a critical effect on the quantum state of the photon and consequently on its quantum-state purity and the photon statistics of the Raman scattering.

We present one- and three-dimensional models of the Raman interaction, corresponding respectively to flat-phase-front (e.g.~waveguided) and free-space propagation of pump and Stokes pulses. We then present experimental data on the degree of photon--matter correlation as a function of pump bandwidth, as measured through second-order coherence of the optical field. Our analysis predicts further correlations due to collection geometry of the broadband Stokes photons, and we confirm experimentally the generation of highly correlated photon--excitation pairs produced when the photons are emitted counter-propagating with the pump. Understanding these correlations and the properties of the joint photon--excitation state, especially in the low gain regime as studied here, is key to controlling Raman emission and enabling new applications for quantum communication, computation, and sensing. We anticipate this work to inform Raman scattering at the quantum level in solid-state systems \cite{Lee11, Lee11_2} as well as atomic vapors \cite{Chaneliere05, Jin18}, and thence on the implementation of quantum protocols such as the Duan-Lukin-Cirac-Zoller protocol \cite{DLCZ} and Raman-based quantum memories \cite{Lee11, Reim11, Matsukevich04, deRiedmatten08, Reim10, Simon10, Kasperczyk15, Wasilewski17, Wasilewski12, Wasilewski17_2}.

\begin{figure*}[t]
	\centering
	\includegraphics[width=\textwidth]{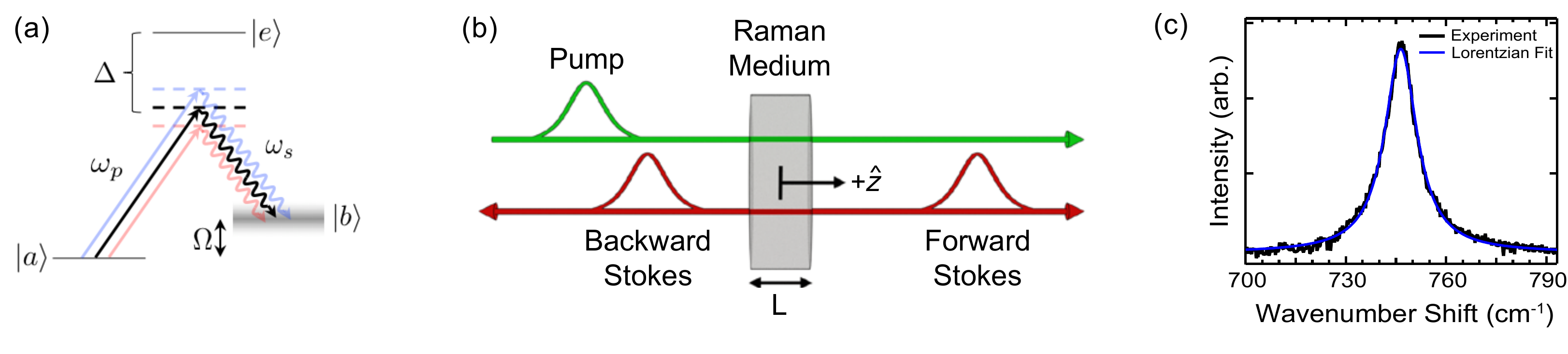}
	\caption{(a) Relevant level structure of the Raman-active medium, where from ground state $\ket{a}$ a pump photon (of angular frequency $\omega_p$) creates a single excitation ($\Omega$) in the medium through far-detuned ($\Delta$) intermediate state $\ket{e}$, leading to collective state $\ket{b}$ and an emitted Stokes photon ($\omega_s$). (b) Schematic of one-dimensional propagation of Stokes photons, emitted forward or backward relative to the pump. (c) Measured gain spectrum of our Raman medium (Al$_2$O$_3$), with a Lorentzian fit. }
	\label{fig:Raman}
\end{figure*}

\section{One-Dimensional Model}\label{1DSection}
The Raman interaction that we consider in this work is shown in the three-level $\Lambda$-system in Fig.~\ref{fig:Raman}(a). Given a laser pump pulse traveling through a Raman-active medium along the $\hat{z}$-axis [Fig.~\ref{fig:Raman}(b)], the Hamiltonian of the system is given by
\begin{equation}\label{eq:Hamiltonian}
    \hat{H}(t) = \hat{H}_{O} + \hat{H}_M + \hat{V}(t),
\end{equation}
where 
\begin{equation}\label{eq:H_O}
\hat{H}_{O}=\int d\omega_s\, \hbar \omega_s \hat{a}^\dag(\omega_s)\hat{a}(\omega_s)    
\end{equation}
is the free Hamiltonian of the Stokes optical field generated in the Raman interaction, and $\hat{a}^\dag(\omega_s)$ ($\hat{a}(\omega_s)$) is the creation (annihilation) operator for a photon of that field with angular frequency $\omega_s$. We assume that the medium's excited states form a bosonic field (Q-field) of collective excitations (CEs)~\cite{raymer_stimulated_1981} with associated creation and annihilation operators $\hat{Q}^\dag(z)$ and $\hat{Q}(z)$, respectively, that obey the commutation relations $[\hat{Q}(z),\hat{Q}(z')]=[\hat{Q}^\dagger(z),\hat{Q}^\dagger(z')]=0$ and $[\hat{Q}(z), \hat{Q}^\dagger(z')]=\delta(z-z')$. $\hat{H}_M$ is the matter Hamiltonian that describes the energy of the Q-field, the environment, and the interaction between the two, which we assume takes form~\cite{scully_zubairy_1997,Rosenau_da_Costa_200}
\begin{equation}\label{eq:H_M}
\begin{split}
\hat{H}_M &= \hbar \Omega_0 \int_{-L/2}^{L/2} dz\, \hat{Q}^\dag(z) \hat{Q}(z)\\
&\hspace{1em}  + \int d\Omega \int_{-L/2}^{L/2} dz\, \hbar \Omega \hat{R}^\dagger(\Omega, z) \hat{R}(\Omega, z)\\
&\hspace{1em}  + \int d\Omega \int_{-L/2}^{L/2} dz\, v(\Omega) \Big [ \hat{R}^\dagger(\Omega, z)\hat{Q}(z)
\\ &\hspace{11.5em} + \hat{R}(\Omega, z)\hat{Q}^\dagger(z) \Big ],
\end{split}
\end{equation}
where $\Omega_0$ is the angular frequency of the Q-field and the environment is treated as a reservoir comprised of a spectral continuum of localized harmonic oscillators with creation (annihilation) operators $\hat{R}^\dagger(\Omega, z)$ ($\hat{R}(\Omega, z)$) for an oscillator with angular frequency $\Omega$ at point $z$, where $v(\Omega)$ is the frequency-dependent coupling between the Q-field and the reservoir, which we take to be real for convenience, and is responsible for the decay of the Q-field and its finite lifetime. The three-wave mixing Raman interaction between the pump, Stokes field, and the medium is given by
\begin{multline}\label{eq:Vt}
    \hspace{-.58em} \hat{V}(t) = \gamma\int d\omega_s\int_{-L/2}^{L/2} dz\,  {E}_p(z,t) \hat{a}^\dag(\omega_s) e^{-i k(\omega_s)z} \hat{Q}^\dag(z)\\ + \text{h.c.},
\end{multline}
where $\gamma$ is a coupling constant dependent on the properties of the Raman medium and the frequency of the Raman emission \cite{Mukamel}, ${E}_p(z, t)$ is the electric field amplitude of the strong pump pulse at time $t$ and point $z$ along the medium, which we treat classically, and $k(\omega_s)$ is the wavevector of the Stokes photon. In Eqs.~(\ref{eq:H_O})-(\ref{eq:Vt}) we consider one-dimensional propagation of the optical fields, which is valid when the interaction medium is a waveguide as well as in various bulk experimental geometries with Fresnel number $\mathscr{F} = A/\lambda L\gg1$, where $A$ is the cross-sectional area of a pencil-shaped beam of wavelength $\lambda$ incident on a Raman medium of length $L$ \cite{Raymer90,raymer_stimulated_1981,Raymer_1985,Raymer:82}. In Section~\ref{StateSec} we consider correlations within photon--CE pairs, in both the forward- and backward-scattering one-dimensional geometries shown in Fig.~\ref{fig:Raman}(b). In Section~\ref{3DSection} we take into account the spatial modes of the optical fields and verify the limit on $\mathscr{F}$ under which this one-dimensional approximation holds, in addition to considering off-axis emission and collection of the Stokes field. 

We begin by examining the medium Hamiltonian in Eq.~(\ref{eq:H_M}); using the procedure in Ref.~\cite{Fano1961}, one can write it in terms of decoupled oscillators~\cite{Rosenau_da_Costa_200}
\begin{equation}\label{eq:H_M_diag}
\hat{H}_M = \int d\Omega \ \hbar\Omega \int_{-L/2}^{L/2}dz \, \hat{B}^\dag(\Omega, z) \hat{B}(\Omega, z),    
\end{equation}
where $\hat{B}(\Omega, z)$ ($\hat{B}^\dagger(\Omega, z)$) is the annihilation (creation) operator associated with a CE (B-field) at point $z$ along the medium with angular frequency $\Omega$, and obeys boson commutation relations. It is given by a linear combination of the Q- and reservoir-fields as $\hat{B}(\Omega,z) =  g(\Omega)\hat{Q}(z) + \int d\Omega^\prime\, h(\Omega, \Omega^\prime)\hat{R}(\Omega^\prime,z)$, where the general solutions for $g(\Omega)$ and $h(\Omega, \Omega^\prime)$ can be found in Ref.~\cite{Rosenau_da_Costa_200}. For this work it is only important to notice that, inversely, we can  express the Q-field operators in terms of B-field operators as~\cite{Rosenau_da_Costa_200}
\begin{subequations}\label{eq:Q}
\begin{align}
&\hat{Q}(z)=\int d\Omega\, g^*(\Omega)\hat{B}(\Omega, z),\\ 
&\hat{Q}^\dag(z)=\int d\Omega\, g(\Omega)\hat{B}^\dag(\Omega, z),    
\end{align}
\end{subequations}
and that $g(\Omega)$ is 
a normalized function ($\int d \Omega\, |g(\Omega)|^2=1$). 
In the case where the coupling in Eq.~(\ref{eq:H_M}) is frequency independent (i.e. $v(\Omega) = v_0$ where $v_0$ is a constant), the Raman gain is homogeneously broadened and 
\begin{equation}
g(\Omega) = \sqrt{\frac{\Gamma/2\pi}{(\Omega-\Omega_0)^2 + (\Gamma/2)^2}}
\end{equation}
takes the form of a Lorentzian lineshape~\cite{Rosenau_da_Costa_200}, where $\Gamma=2\pi|v_0|^2$ is the full width at half maximum bandwidth of the Raman-gain spectral intensity. In the eigenbasis of $\hat{H}_M$ [Eq.~(\ref{eq:H_M_diag})], the interaction term in Eq.~(\ref{eq:Vt}) is 
written as
\begin{equation}\label{eq:Vt_2}
\begin{split}
\hspace{-.58em} \hat{V}(t) &= \gamma \int d\Omega d\omega_s \int_{-L/2}^{L/2} dz \, {E}_p(z, t)  \hat{a}^\dag(\omega_s)e^{-i k(\omega_s)z} \\ &\hspace{10em} \times g(\Omega)\hat{B}^\dag(\Omega,z) + \text{h.c.},
\end{split}    
\end{equation}
\noindent which couples the Stokes field to a spectral continuum of distinct oscillators (B-field), with coupling amplitude $\gamma g(\Omega)$. In effect, Eq.~(\ref{eq:Vt_2}) 
mathematically treats homogeneous broadening as inhomogeneous broadening with a Lorentzian lineshape; such equivalence has been found empirically in 
the analysis of the statistical properties of the optical field operators as derived in the fluctuation-dissipation approach (that is, writing the Heisenberg equations of motion with the Q-field operators) in Ref.~\cite{Rzazewski1989}. 
We note, however, that unlike the case of inhomogeneous broadening where the lineshape is dictated by the (Gaussian) distribution of the density of states, here the spectral distribution of the density of states is uniform, while the coupling strength ($|\gamma g(\Omega)|^2$) is responsible for the lineshape.

Transforming Eq.~(\ref{eq:Vt_2}) into the interaction picture, after the rotating wave approximation we arrive at the interaction Hamiltonian
\begin{equation}\begin{split}\label{eq:H_I}
    \hspace{-0.7em}\hat{H}_{I}(t) &= \gamma \int d\omega_p d\omega_s d\Omega
        \int_{-L/2}^{L/2} dz\, \bigg[ \mathcal{E}(\omega_p)g(\Omega) \\ &\hspace{1.2em} e^{i[k(\omega_p) - k(\omega_s)] z} e^{-i\Delta\omega t} \hat{a}^\dag(\omega_s) \hat{B}^\dag(\Omega, z)\bigg] + \text{h.c.},
\end{split}
\end{equation}
where we have written the classical pump field in the spectral domain as $E_p(z, t)=\int d\omega_p\, \mathcal{E}(\omega_p) e^{i[k(\omega_p)z - \omega_p t]} + \textrm{h.c.}$ with the spectral amplitude $\mathcal{E}(\omega_p)$ and wavenumber $k(\omega_p)$. The frequency mismatch of the three fields is $\Delta\omega = \omega_p - \omega_s - \Omega$, and for simplicity we assume the pump and Stokes modes 
have the same dispersion relation; it is straightforward to expand our treatment when this is not the case.


\section{Photon--CE Pair State}\label{StateSec}

\begin{figure*}[t]
	\centering
	\includegraphics[width=\textwidth]{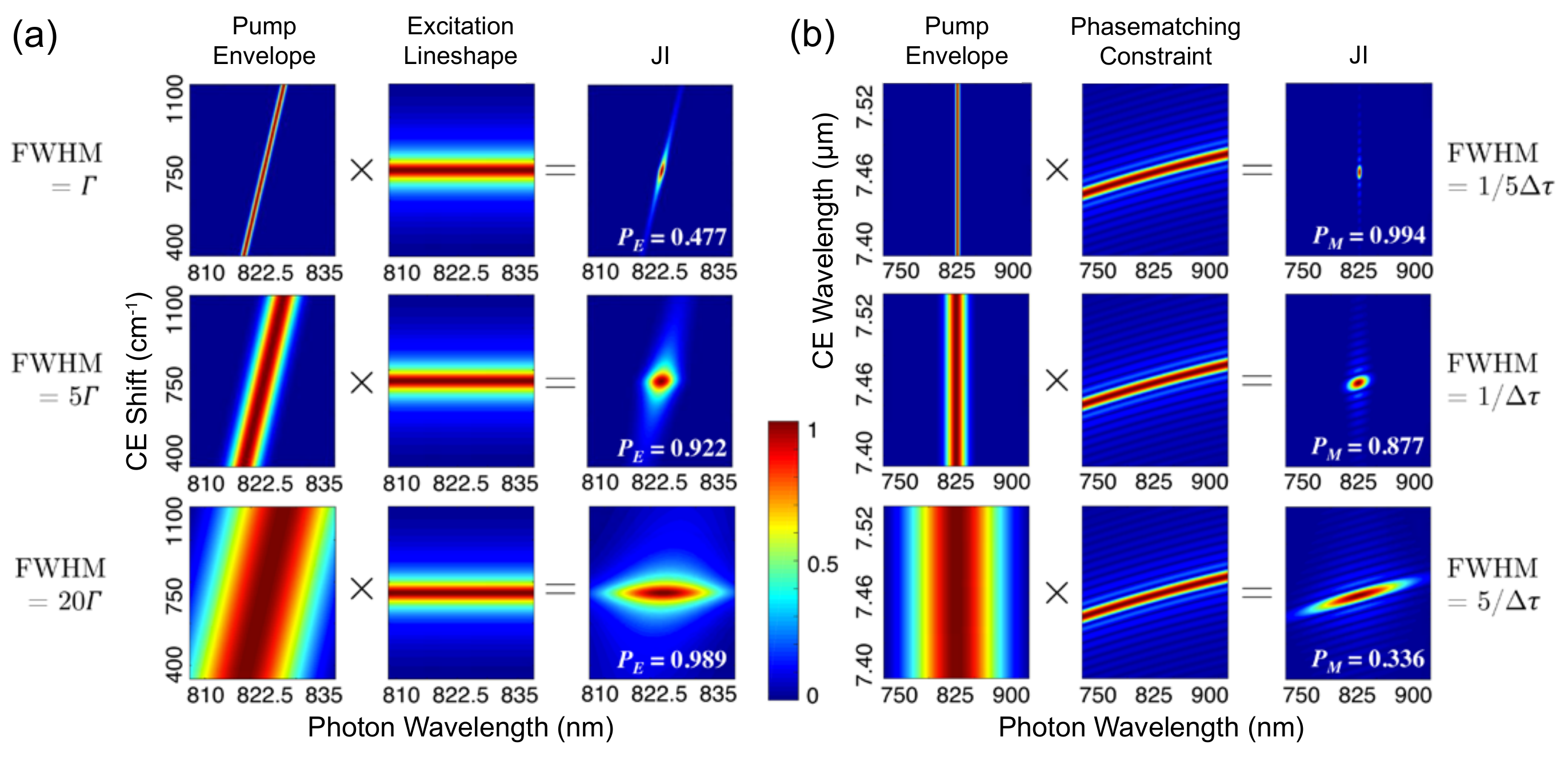}
	\caption{Joint intensity (JI)  of photon--CE pairs in the one-dimensional regime for (a) fixed position, showing the effect of excitation linewidth, and (b) fixed collective excitation (CE) frequency, showing the effect of chromatic dispersion; FWHM is the pump spectral full width at half maximum, $\Gamma$ the linewidth of the relevant excitation, $\Delta\tau$ the group delay between Stokes and pump pulses for a bulk Al$_2$O$_3$ Raman medium, and $P_E$ and $P_M$ the frequency (energy)- and momentum-state purities corresponding to each joint intensity, respectively.
	}\label{fig:JSIs}
\end{figure*}

We assume that prior to the pump pulse entering the medium the Stokes and CE fields are both in their respective vacuum states, which we write in the combined Stokes--CE system as $\vac$. In this work we restrict our discussion to the low-gain limit, assuming that the interaction is weak and perturbative expansion of the resulting state is allowed. Once the interaction ceases, the lowest order non-vacuum state of this system~\cite{Grice97} $\ket{\Psi} = \mathcal{N}\int_{-\infty}^{+\infty} dt \, \hat{H}_I(t) \vac$ describes a joint quantum state of a single Stokes photon and collective excitation, where $\mathcal{N}$ is a normalization factor.
In the one-dimensional case, the resulting photon--CE pair state is $\ket{\Psi} = \mathcal{N}\int d\omega_s d\Omega\int_{-L/2}^{L/2}dz\,  f_\textrm{1D}(\omega_s,\Omega,z) \ket{\omega_s;\Omega, z}$, where the joint amplitude (JA) for the photon--CE pair is given by
\begin{align}
\label{eq:f1D}
f_\textrm{1D}(\omega_s,\Omega,z) &= \mathcal{E}(\omega_s+\Omega)g(\Omega)e^{i[k(\omega_s + \Omega) - k(\omega_s)]z}\\
\begin{split}\label{jointAmpApprox}
\hspace{0.3em}\approx \mathcal{E}(\omega_p^0+\nu + \delta)g(\Omega^0+\delta) e^{i\left(\Delta\tau/L\right)\nu z} \\ \hspace{+0em} \times \,  e^{i\beta_p\delta z} e^{i\left[k(\omega_p^0) - k(\omega_s^0)\right]z},
\end{split}
\end{align}
\noindent and $\ket{\omega_s;\Omega, z}=\hat{a}_s^\dagger(\omega_s) \hat{B}^\dagger(\Omega,z)\vac$ represents a photon--CE pair state with Stokes photon angular frequency $\omega_s$ and CE with angular frequency $\Omega$ created at point $z$ along the interaction medium. Here we define the center frequency of the pumping light $\omega_p^0$, which is set by experiment, and the center frequency of the CE mode $\Omega^0$, which is determined by the properties of the Raman medium and in turn defines the center frequency of the Stokes light $\omega_s^0= \omega_p^0 - \Omega^0$ through energy conservation. The variations about the center frequencies $\nu = \omega_s - \omega_s^0$, $\delta = \Omega - \Omega^0$, and linear expansions $k(\omega_s+\Omega) - k(\omega_s^0+\Omega^0) \approx \beta_p(\nu+\delta)$ and $k(\omega_s) - k(\omega_s^0) \approx \beta_s\nu$, where $\beta_{p(s)}=\partial k/\partial \omega|_{\omega_{p(s)}^0}$ is the inverse group velocity of the pump (Stokes) pulse lead to the approximate form in Eq.~(\ref{jointAmpApprox}),
\noindent where $\Delta\tau = (\beta_p-\beta_s)L$ is the group delay between pump and Stokes pulses acquired during propagation in a dispersive medium. 

We have decomposed Eq.~(\ref{jointAmpApprox}) into terms with different physical roles: the $\mathcal{E}(\omega_p^0 + \nu + \delta)$ term manifests energy conservation via the pump spectral envelope, the lineshape $g(\Omega^0 + \delta)$ puts constraints on the value of the CE energy, $e^{i\left(\Delta\tau/L\right)\nu z}$ is responsible for correlations between the point at which the CE is created and the arrival time of the Stokes photon~\cite{Wasilewski_2006}, and the $e^{i\beta_p\delta z}$ term describes correlations due to the fact that a CE created at one point in the medium has evolved and decayed more than those created later. This latter term raises correlations between the position and energy of the CE and thus affects the CE internal state, but has no effect on the state of the Stokes photon. The term $e^{i\left[k(\omega_p^0) - k(\omega_s^0)\right]z}$ represents a global phase accumulation and does not possess any correlations. 

With the Fourier transform of the CE creation operator $\hat{b}^\dagger(\Omega,k_{\mathrm{CE}}) = (2\pi)^{-1}\int dz\, \hat{B}^\dagger(\Omega,z)e^{ik_{\mathrm{CE}} z}$, where $k_{\mathrm{CE}} = k_{\mathrm{CE}}^0 + \kappa$ is the CE wavevector with variation $\kappa$ about its center $k_{\mathrm{CE}}^0=k(\omega_p^0) - k(\omega_s^0)$, the $k$-space JA is given by
\begin{align}
\label{jointAmpkSpace}\notag \hspace{-0.5em} \widetilde{f}_\textrm{1D}(\omega_s,\Omega,k_{\mathrm{CE}}) & \\ &\hspace{-6.5em}=\mathcal{E}(\omega_s + \Omega)g(\Omega)\textrm{sinc}\left[\frac{L}{2}\big(k(\omega_s+\Omega) - k(\omega_s) - k_{\mathrm{CE}}\big)\right]\\
 \label{jointAmpkSpaceApprox} &\hspace{-6.5em}\approx \mathcal{E}(\omega_p^0 + \nu + \delta)g(\Omega^0 + \delta)\textrm{sinc}\bigg[\frac{\Delta\tau}{2}\nu + \frac{L}{2}\big(\beta_p\delta - \kappa \big)\bigg].
\end{align}
The joint amplitudes in Eqs.~(\ref{eq:f1D})-(\ref{jointAmpkSpaceApprox}) capture the spectral and momentum correlations between spontaneous Stokes photon and CE in one dimension, including those arising from the CE linewidth and group velocity dispersion in the medium.  

The quantum state of the Stokes photon created in this interaction is given by the reduced density matrix
\begin{align}\label{eq:rho_s_1d}
    \notag\hat{\rho}_s&= \text{Tr}_{\mathrm{CE}}\ket{\Psi}\bra{\Psi} \\ \notag &= \mathcal{N}^2\int d\omega_s d\omega_s'd\Omega dz \, f_\text{1D}(\omega_s,\Omega,z)f_\text{1D}^*(\omega_s',\Omega,z)\\ &\hspace{11em} \times \ket{\omega_s;\Omega,z}\bra{\omega_s';\Omega,z},
\end{align}
where Tr$_{\mathrm{CE}}$ represents the partial trace over the CE degrees of freedom, $\Omega$ and $z$. The quantum state purity of the Stokes photon $P = \text{Tr}\hat{\rho}_s^2$ amounts to the degree to which the photon and CE are in pure rather than mixed states, and is a critical figure of merit in quantum protocols that rely on two-photon interference~\cite{HOM}. In particular, the photon--CE pair state that leads to unit purity of the Stokes photon is the  factorable  state, where the JA can be written as independent functions of the Stokes and CE degrees of freedom: $f(\omega_s,\Omega,z) = f_s(\omega_s)f_{\mathrm{CE}}(\Omega,z)$. Conversely, when the photon and CE are spectrally entangled, $f(\omega_s,\Omega,z)$ is not factorable, $P <1$, and the photon and CE are individually in mixed states.

In general, all three degrees of freedom of the photon--CE pair are entangled. In order to characterize this entanglement, we consider correlations between the photon frequency and each degree of freedom of the CE in turn. These two forms of entanglement arise mainly from two separate physical effects, which we explore in the following subsections. To enumerate these correlations, unless otherwise stated we consider a single-crystal, c-axis, bulk sapphire (Al$_2$O$_3$) Raman medium of length $L = 8$ mm, with measured 746.6 cm$^{-1}$ Raman shift and Lorentzian lineshape with full width at half maximum (FWHM) $\Gamma = 11.0$ cm$^{-1}$ [see Fig.~\ref{fig:Raman}(c)] corresponding to ($2\bar{1}\bar{1}0$)$E_g$ optical phonon creation in the medium \cite{Ashkin68,Pezzotti15}. We consider pump pulses centered at 775 nm and approximate chromatic dispersion in the bulk with the Sellmeier equation of Ref.~\cite{Dodge86}.

\subsection{Effect of Excitation Linewidth}

To isolate the effect of a finite CE linewidth on spectral correlations between photon and excitation, which exist on the local level of the CE 
(i.e. for each fixed location $z = z^0$)
, we write the components of the JA that capture these energy correlations as
\begin{equation}\label{eq:fE_1D}
f^E_\textrm{1D}(\omega_s,\Omega) = \mathcal{E}(\omega_s+\Omega)g(\Omega).
\end{equation}
Figure~\ref{fig:JSIs}(a) shows the components of the photon--CE joint intensities (JIs) $\abs{f^E_\textrm{1D}(\omega_s,\Omega)}^2$ for pump pulses around the intermediate regime FWHM $\sim\Gamma$, where FWHM is the spectral intensity full width at half maximum of a Gaussian pump envelope. 
We plot these JIs with respect to photon wavelength $\lambda_s = 2\pi c/\omega_s$ and CE shift $\nu_\text{CE} = \Omega/(2\pi c)$, where $c$ is the speed of light. The results of Fig.~\ref{fig:JSIs}(a) show an increase in pair correlations with decreasing pump spectral width. Physically this indicates energy entanglement between photon and CE: in the limit of a monochromatic pump, the linewidth of the CE allows for a distribution of Stokes photons in frequency, with each frequency entangled with an excitation through energy conservation. For larger bandwidths or narrower CE linewidths this entanglement is diminished. As a figure of merit, we also include in Fig.~\ref{fig:JSIs}(a) the energy state purity $P_E = \text{Tr}\hat{\rho}_{s,E}^2$ (where $\hat{\rho}_{s,E}$ is the reduced energy state density matrix of the Stokes photon, given by the trace of $\hat{\rho}_s$ only over CE frequency $\Omega$ at fixed $z^0$) corresponding to each JI and calculated photon--CE state. In the absence of further momentum state correlations, $P_E=P$ is the quantum state purity of the Stokes photon.

\subsection{Effect of Chromatic Dispersion}

As derived in previous work in the time-domain \cite{Wasilewski_2006}, chromatic dispersion leads to entanglement between the Stokes frequency and the location or momentum of the CE in the Raman medium. 
Figure~\ref{fig:JSIs}(b) shows the JI components in $k$-space for pump pulse durations varying about the group delay between Stokes and pump pulses,  which is $\Delta\tau \approx 32$ fs for the medium considered here. To isolate the correlations between photon frequency and CE momentum ($k_{\mathrm{CE}}$), we write the JA at fixed CE frequency
\begin{multline}\label{eq:fM_1D}
    \hspace{-0.5em}\tilde{f}^M_\text{1D}(\omega_s,k_{\mathrm{CE}}) \\ = \mathcal{E}(\omega_s+\Omega^0)\text{sinc}\left[\frac{L}{2}\left(k(\omega_s+\Omega^0)-k(\omega_s)-k_{\mathrm{CE}}\right)\right].
\end{multline}
In Fig.~\ref{fig:JSIs}(b) $\abs{f^M_\text{1D}(\omega_s,k_{\mathrm{CE}})}^2$ is plotted against photon and CE wavelength ($2\pi/k_{\mathrm{CE}}$), showing the effect of chromatic dispersion in the absence of those correlations considered in Fig.~\ref{fig:JSIs}(a). For a given interaction length, the group-delay between pump and Stokes pulses leads to momentum correlations between photon and CE, due to the temporal walkoff between pulses that serves to distinguish the spatial location of photon--CE pair creation. For larger pump bandwidths 
(shorter 
coherence-lengths) 
the distinguishability of Stokes pulses increases, increasing the photon--CE correlations. Conversely, for smaller medium lengths the accumulated group delay between Stokes and pump pulses and the resulting correlations decrease. We include in Fig.~\ref{fig:JSIs}(b) the momentum state purity ($P_M = \text{Tr}\hat{\rho}_{s,M}^2$, where $\hat{\rho}_{s,M}$ is given by the trace of $\hat{\rho}_{s}$ only over CE position $z$ at fixed $\Omega^0$) corresponding to each pair state.

\begin{figure}[t]
	\centering
	\includegraphics[width=\linewidth]{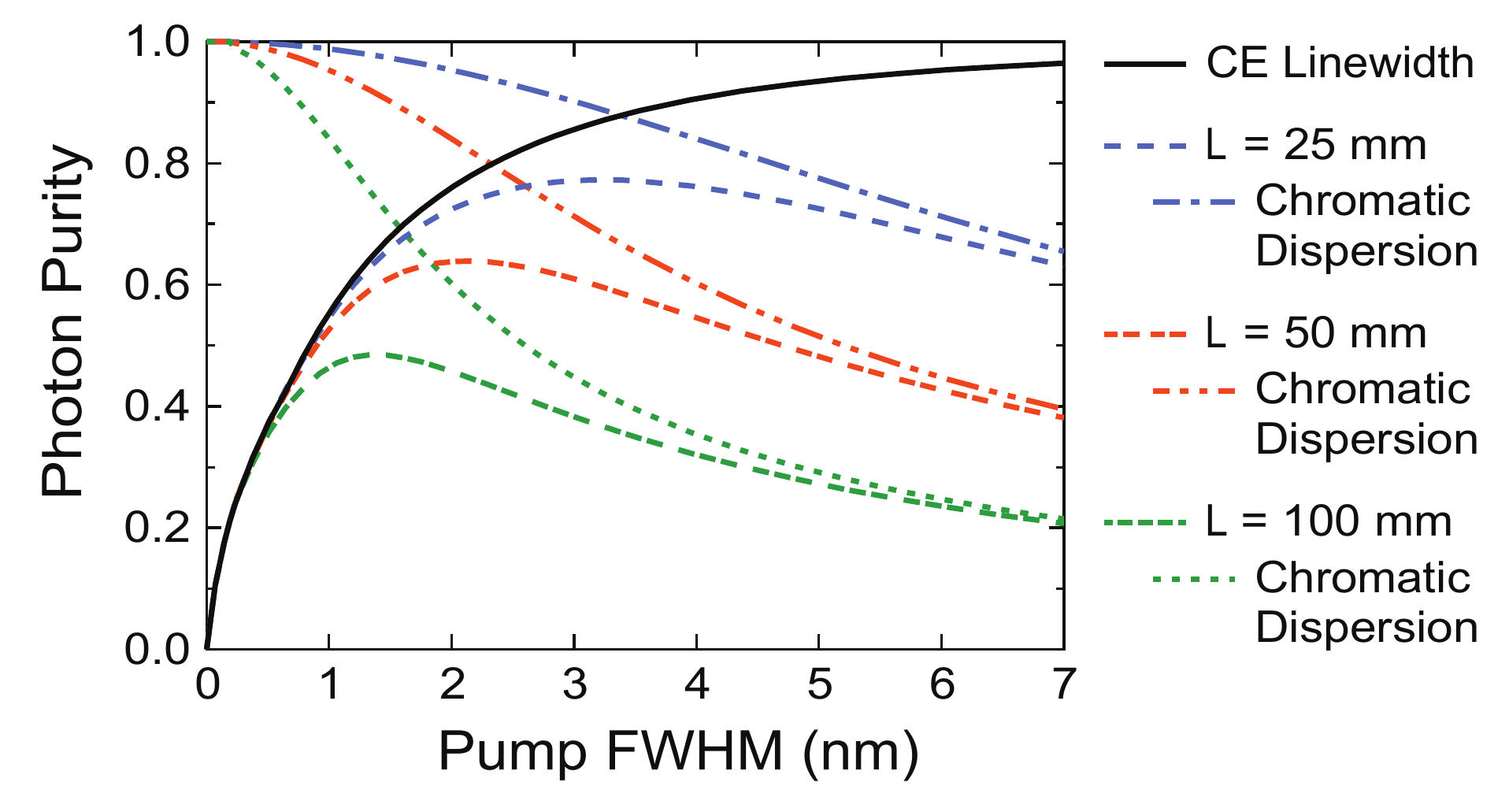}
	\caption{Stokes photon purities calculated within a one-dimensional model (see text) as a function of pumping spectral full width at half maximum (FWHM) for varying interaction lengths of a bulk Al$_2$O$_3$ Raman medium. Effects on the photon purity due to the collective excitation (CE) linewidth (solid) and chromatic dispersion (dotted) are isolated for a given length, showing their contribution to each convolution (dash-dot) representing the total photon purity.}
	\label{fig:P_v_BW_L}
\end{figure}

\vspace{1em}

The competing effects of excitation linewidth and chromatic dispersion in general lead to a maximum Stokes photon purity (minimum photon--CE entanglement) at finite pump bandwidth. Within this one-dimensional model, the magnitude of this maximum purity along with the corresponding pump settings are dependent only on three parameters: the dispersion relation in the medium, the linewidth of the excitation, and the interaction length. In practice, while the dispersion relation and linewidth of an excitation are not easily modified parameters for a given Raman medium, the interaction length is. In Fig.~\ref{fig:P_v_BW_L} then, we show the dependence of the spontaneous Stokes photon purity on pump spectral bandwidth for varying interaction length, along with the isolated effects of finite excitation linewidth and chromatic dispersion, to show the general behavior of the photon purity.

\subsection{Backward Collection}

As spontaneous Stokes scattering is generally emitted over solid angle $4\pi$, we extend our one-dimensional treatment to backward-emission of spontaneous Stokes photons [Fig.~\ref{fig:Raman}(b)]. We find the pair state joint amplitude:
\begin{equation}
\label{f1Dbackward}f_\textrm{1D}^\leftarrow(\omega_s,\Omega,z) = \mathcal{E}(\omega_s+\Omega)g(\Omega)e^{i[k(\omega_s+\Omega)+k(\omega_s)]z}.
\end{equation}

In general this modification serves to increase photon--CE correlations via temporal walkoff by an argument analogous to that of chromatic dispersion in the previous subsection, though they are, to be clear, independent effects: Even in the absence of dispersion, the temporal delay between Stokes photons generated at the input versus the output face of the medium under consideration is $\Delta\tau^\leftarrow \approx 26$~ps. This timing information serves to distinguish photon--CE pairs and decrease the state purity of the photon, and normally 
has a significantly stronger effect on the photon purity than chromatic dispersion. Here we note that whereas for Raman-active atomic vapors the effect of excitation linewidth and chromatic dispersion may be negligible, for the same media collection in the backwards direction~\cite{Jin18,Yu19} can strongly affect the correlations between broadband photon and excitation.

\section{Three-Dimensional Model:\\ Free-Space Propagation}\label{3DSection}

\begin{figure}[t]
	\centering
	\includegraphics[width=\linewidth]{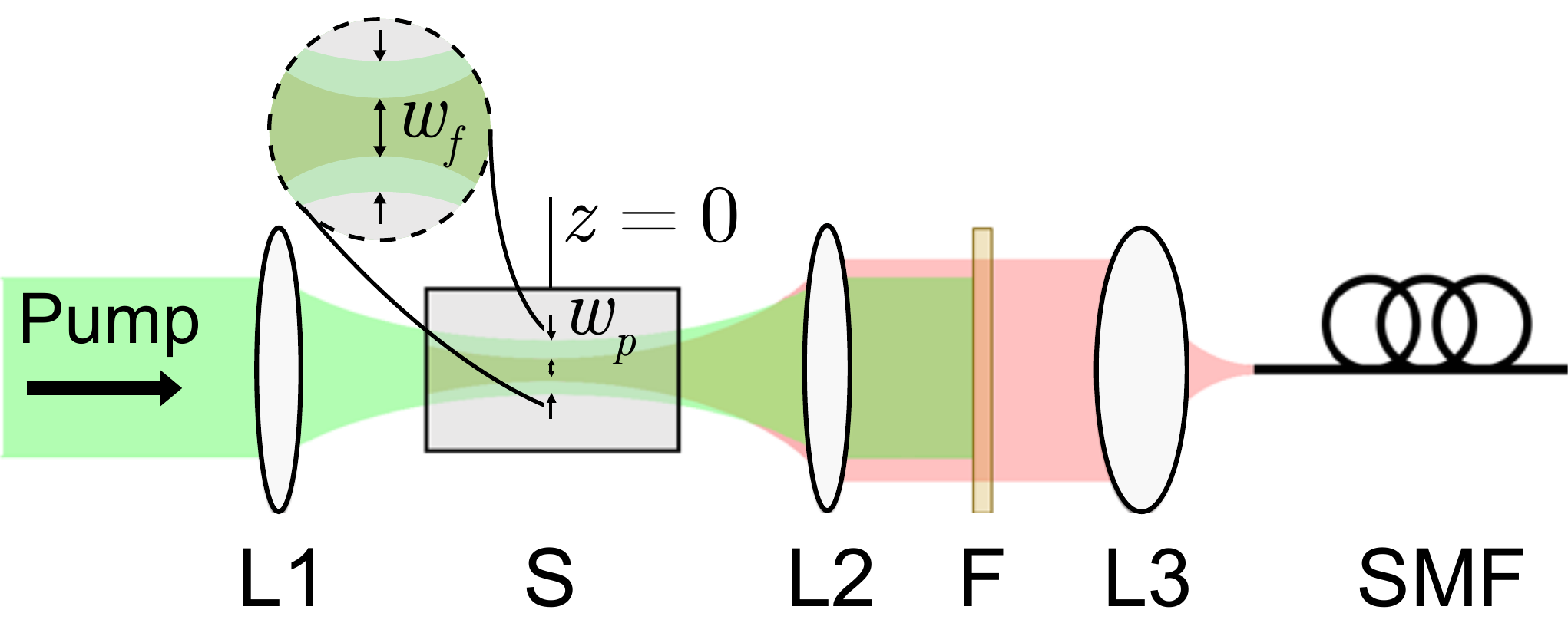}
	\caption{Configuration assumed for collinear three-dimensional calculations. A pump beam (green) is focused by a lens (L1) into the Raman medium (S). Stokes photons are generated, collected by a lens (L2), pass through a spectral filter (F) that removes the pump light, and are coupled by a coupling lens (L3) to single mode fiber (SMF), projecting the photons onto an approximately Gaussian collection mode (red).}
	\label{fig:3Dcollection}
\end{figure}

In the case of bulk optics and free-space propagation of the optical fields, we consider a $\text{TEM}_{00}$ Gaussian pump beam focused into the Raman medium that reaches its minimal beam waist radius $w_p$ at the center of the medium (see Fig.~\ref{fig:3Dcollection}). While the subsequently generated Stokes field is generally emitted in all directions, here we consider the quantum state of only the fraction of photons that are collected by a lens and coupled into a single-mode fiber. In this three-dimensional case we cannot neglect correlations in the transverse degrees of freedom of photon and CE \cite{DCZ_3D,Mostowski84}. To include these correlations we rewrite the  pump, 
CE, and Stokes fields with the additional cylindrically symmetric degree of freedom $\boldsymbol{\rho}=(x,y)$: The Q-field operators take the form 
\begin{equation}\label{eq:Q_rho}
\hat{Q}^\dagger(\boldsymbol{\rho},z) = \int d\Omega d^2\mathbf{q}_{\mathrm{CE}}\, g(\Omega) e^{-i\mathbf{q}_{\mathrm{CE}}\cdot\boldsymbol{\rho}} \hat{B}^\dagger(\Omega,\boldsymbol{q}_{\mathrm{CE}},z), 
\end{equation}
where the CE field with creation operator  $\hat{B}^\dagger(\Omega,\boldsymbol{q}_{\mathrm{CE}},z)$ now also includes the CE transverse wavevector ($\boldsymbol{q}_{\mathrm{CE}}$) as an additional degree of freedom. We assume that the spatial mode supported by the single-mode fiber (into which the Stokes photons are collected) can also be well approximated by a $\textrm{TEM}_{00}$ Gaussian mode such that the Stokes collection configuration of Fig.~\ref{fig:3Dcollection} projects the optical field onto the Gaussian state $\ket{u_f(\omega_{s})} = \hat{A}^{\dag}(\omega_s)\vac$ with beam waist $w_f$, which we assume to also occur at the center of the Raman medium, where 
\begin{equation}\label{eq:A_s}
    \hat{A}^{\dag}(\omega_s) = (4\pi/w_f^2)\int d^2\mathbf{q}_s \,  e^{-w_f^2\vert\mathbf{q}_s\vert^2/4} \hat{a}^\dagger_{\mathbf{q}_s}(\omega_s)
\end{equation}
and $\hat{a}^\dagger_{\mathbf{q}_s}(\omega_s)$ is the creation operator of a Stokes photon with transverse wavevector $\mathbf{q}_s$ and angular frequency $\omega_s$. 
The resulting projected state is then given by $\label{3DStateProjected} \ket{\Psi}_\textrm{3D}^\text{proj.} = \mathcal{N}_\text{3D}\int d\omega_s d\Omega d^2\mathbf{q}_{\mathrm{CE}} \int_{-L/2}^{L/2}dz \,  f_\textrm{3D}(\omega_s, \Omega, z, \mathbf{q}_{\mathrm{CE}})  \hat{A}_s^\dagger(\omega_s)$ $\hat{B}^\dagger(\Omega,\mathbf{q}_{\mathrm{CE}},z)\vac$ (for a more detailed calculation, see Appendix). Here $\mathcal{N}_\text{3D}$ is the appropriate normalization factor and we find the three-dimensional JA can be expressed in terms of the one-dimensional JA [Eq.~(\ref{eq:f1D})] as
\begin{equation}\label{eq:f_3D}
    f_\textrm{3D}(\omega_s, \Omega, z, \mathbf{q}_{\mathrm{CE}}) = \beta(\mathbf{q}_{\mathrm{CE}},z) f_\textrm{1D}(\omega_s, \Omega, z),\\
\end{equation}
where
\begin{equation}
\label{eq:beta}
	\beta(\mathbf{q}_{\mathrm{CE}},z) = \frac{\exp\left[-i\frac{C_p(z)C^*_s(z)}{2(C_p(z)-C^*_s(z))}\vert\mathbf{q}_{\mathrm{CE}}\vert^2\right]}{C_p(z)-C^*_s(z)}
\end{equation}
	
\noindent and $C_p(z) = (z+i z_{R,p})/k(\omega_p^0)$ and $C_s(z) = (z+i z_{R,f})/k(\omega_s^0)$ for pump and fiber collection modes with Rayleigh ranges $z_{R,p}=k(\omega^0_p)w_p^2/2$ and $z_{R,f}=k(\omega^0_f)w_f^2/2$, respectively. We ignore the slow spectral dependence of $\beta(\mathbf{q}_{\mathrm{CE}},z)$ in considering only central wavevectors $k(\omega_s^0)$ and $k(\omega_p^0)=k(\omega_s^0+\Omega^0)$. We define the Fresnel numbers of the pump and fiber modes in terms of their respective Rayleigh ranges 
as $\mathscr{F}_p= 2z_{R,p}/L$ and  $\mathscr{F}_f = 2z_{R,f}/L$, respectively. As expected, for interaction lengths much smaller than the Rayleigh ranges, i.e.  $\mathscr{F}_p\textrm{, }\mathscr{F}_f \gg 1$, the correction in Eq.~(\ref{eq:beta}) reduces to a constant and the one-dimensional calculations hold. In the three-dimensional case, the reduced density matrix of the Stokes photon is then given by
\begin{multline}\label{eq:denmatrix_3d}
\hspace{-1em} \hat{\rho}_s = \mathcal{N}_\text{3D}^2\int d\omega_s d\omega_s' d\Omega dz \, \alpha(z) f_\text{1D}(\omega_s,\Omega,z) f^*_\text{1D}(\omega_s',\Omega,z)\\  \times \ket{\omega_s;\Omega,z}\bra{\omega_s';\Omega,z}, 
\end{multline}
\noindent where
\begin{align}
\notag \label{alpha}\alpha(z) &= \int d^2\mathbf{q}_{\mathrm{CE}}\,  \abs{\beta(\mathbf{q}_{\mathrm{CE}},z)}^2 \\ &= \frac{8 \pi^3}{w_p^2 w_f^2}\left(\frac{z^2 + z_{R,f}^2}{\left[w_f k(\omega^0_s)\right]^2} + \frac{z^2 + z_{R,p}^2}{\left[w_p k(\omega^0_p)\right]^2} \right)^{-1}
\end{align}
is a Lorentzian function along $z$ that manifests an effective apodization of the interaction length and therefore has the effect of decreasing correlations between the Stokes photon and the spatial (or momentum) degree of freedom of the CE (in comparison to the 1D case).

\subsection{Off-Axis Collection of Stokes Photons}\label{offaxisSection}

We now generalize the above treatment to include off-axis collection of Stokes photons at angle $\varphi$ from the $z$-axis within the free-space model. We assume that the dispersion relation is independent of propagation angle, a condition that is  satisfied for isotropic media such as atomic vapors, or for uniaxial crystalline media with pump and Stokes polarizations along the ordinary axis. It is straightforward to include emission modes with different dispersion relations when this condition is not met. We further assume that the collection and pumping modes share a focal point. Under these assumptions, the photon--CE JA is given by Eq.~(\ref{eq:f_3D}) with a generalized form of Eq.~(\ref{eq:beta}):
\begin{align}\label{eq:beta_phi}
    \notag&\beta(\mathbf{q}_{\mathrm{CE}},z,\varphi) = \\ 
    \notag&\exp\left[ i\frac{ C_p(z)}{2} \left(\abs{\mathbf{q}_{\mathrm{CE}}}^2 - k(\omega_s^0)^2 \sin^2\varphi - 2 k(\omega_s^0) q_{\mathrm{CE}}^y \sin\varphi \right) \right]\\ 
    &\hspace{2em} \times \frac{\exp\left[i\frac{(C_p(z) \cos\varphi (q_{\mathrm{CE}}^y - k(\omega_s^0) \sin\varphi ) + z \sin\varphi)^2}{2  \left(C_s^{'*}(z) - C_p(z) \left(2 \cos ^2\varphi - 1 \right)\right)}\right]}{\sqrt{C_s^{'*}(z) - C_p(z)(2\cos^2\varphi -1)}} \\
    \notag&\hspace{4em} \times \frac{\exp\left[i\frac{\left(C_p(z) q^{x }_{\mathrm{CE}}\right)^2}{2\left(C_s^{'*}(z) -  C_p(z)\cos^2\varphi\right)}  \right]}{\sqrt{C_s^{'*}(z) -  C_p(z)\cos^2\varphi}},
\end{align}

\noindent where $C'_s(z) = (z\cos\varphi+i z_{R,f})/k(\omega_s^0)$ and $q^{x }_{\mathrm{CE}}$ ($q^{y }_{\mathrm{CE}}$) is the transverse momentum component of the CE along the $x$- ($y$-) axis. Similarly, the Stokes photon density matrix takes the same form as Eq.~(\ref{eq:denmatrix_3d}) with the generalized apodization function [Eq.~(\ref{alpha})]
\begin{equation}\begin{split}
	\label{3Doffaxis_alpha}\hspace{0em}\alpha(z,\varphi) &=
	\exp \left[-\frac{2 z^2 \sin ^2\varphi}{w_f^2 + \left(\varpi_f^2(z) + \varpi_p^2(z) + w_p^2\right) \cos^2\varphi}\right]\\
	&\hspace{1em} \times\left\{\left[(\varpi_f^2(z)+\varpi_p^2(z) + w_p^2) \cos^2\varphi +w_f^2\right]\right.\\ 
	&\hspace{3em} \left. \left[\varpi_f^2(z)\cos^2\varphi+\varpi_p^2(z)+w_p^2+w_f^2\right]\right\}^{-1/2},
\end{split}\end{equation}
\noindent where $\varpi_f(z) = w_f z/z_{R,f}$ and $\varpi_p(z) = w_p z/z_{R,p}.$

\begin{figure}[t]
	\centering
	\includegraphics[width=\linewidth]{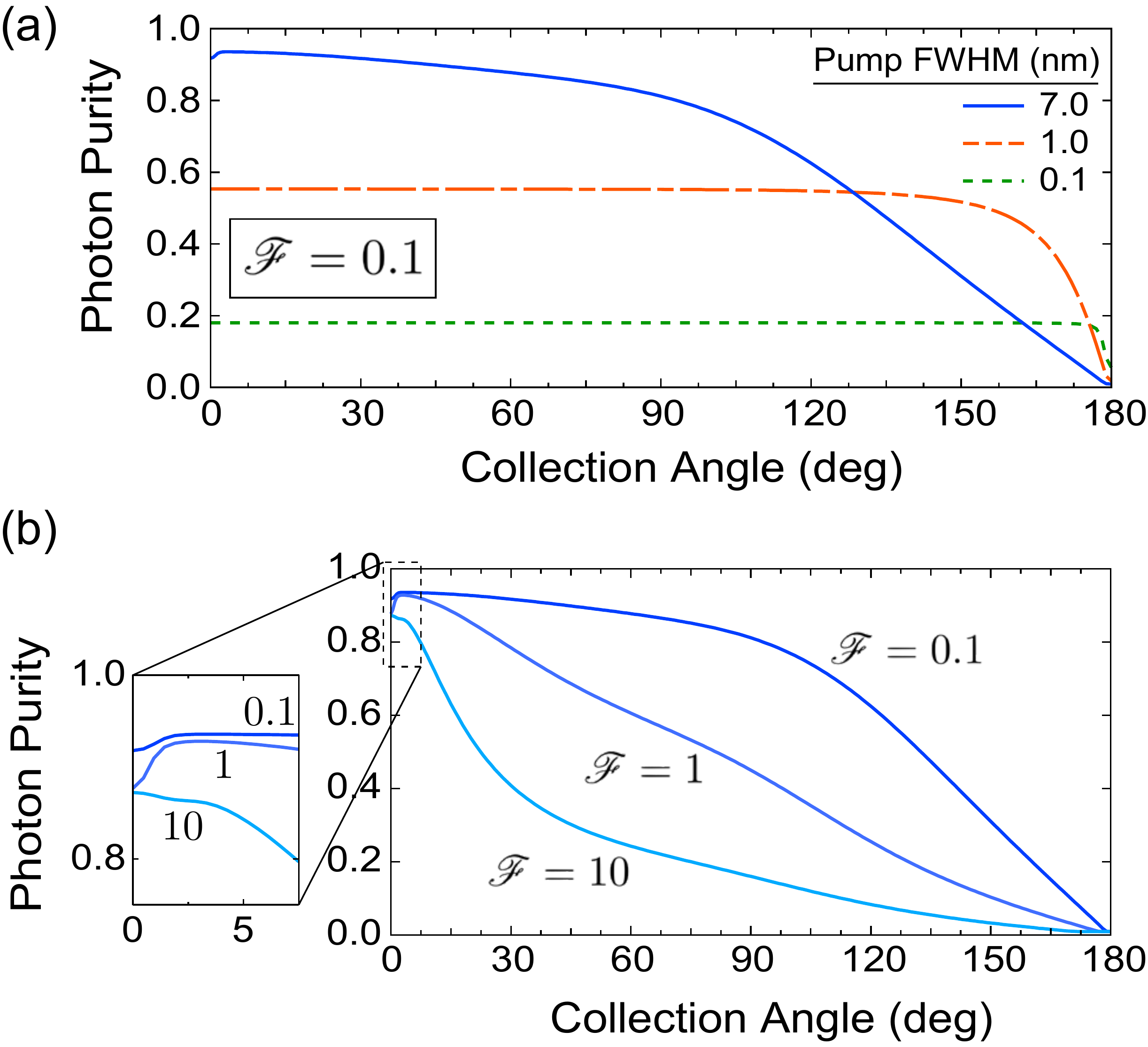}
	\caption{Stokes photon purity (a) as a function of collection angle for varying pump bandwidth at Fresnel number $\mathscr{F}=0.1$, and (b) for varying Fresnel number at fixed pump $\textrm{FWHM}= 7$ nm. All calculations are performed within a three-dimensional model (see text).}
	\label{offaxis_fig}
\end{figure}

Equation~(\ref{3Doffaxis_alpha}) has a Lorentzian form in the co-propagating case ($\varphi=0$) and approaches a Gaussian as $\varphi$ approaches 90$^\circ$. The FWHM of the apodization function---the effective length from which photons are collected---decreases considerably with increasing collection angle up to collection perpendicular to the pump ($\varphi = 90^\circ$), after which it increases symmetrically until the counter-propagating case ($\varphi = 180^\circ$) when again the form is Lorentzian (see Appendix). In general, for a fixed medium length and collection angle, a more tightly focused pump beam will generate a narrower apodization function. The effect of this apodization function on the photon purity is shown in Fig.~\ref{offaxis_fig}(a) as a function of collection angle for varying pump bandwidths, plotted for Fresnel number $\mathscr{F}=\mathscr{F}_p=\mathscr{F}_f=0.1$. We note that in this case, for tightly focused beams, the photon purity is robust to small changes in collection angle about $\varphi=0$. In Fig.~\ref{offaxis_fig}(b) however, we plot the photon purity for varying Fresnel number at fixed pump bandwidth FWHM = 7 nm, and note that for loosely focused beams the purity becomes more sensitive to changes in collection angle. 
For some media (including bulk Al$_2$O$_3$ presented here) and Fresnel numbers, this interaction length apodization can lead to a maximal photon state purity at nonzero collection angle, as shown in the inset of Fig. \ref{offaxis_fig}(b).

\section{Experimental Results}

\begin{figure*}[t]
	\centering
	\includegraphics[width=\linewidth]{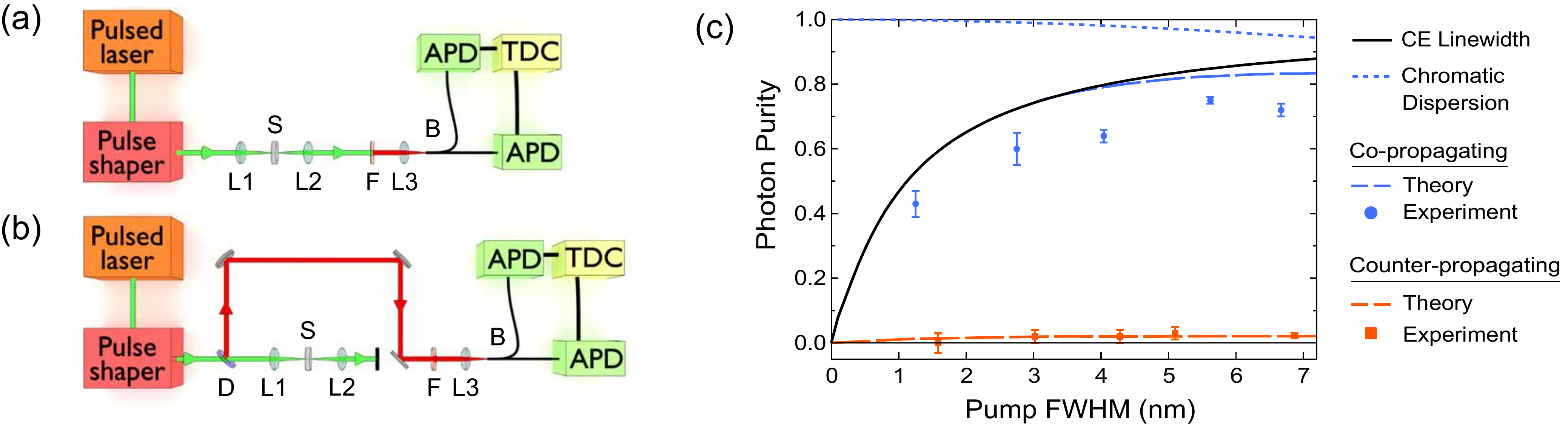}
	\caption{Experimental diagrams of (a) co- and (b) counter-propagating collection of Stokes photons; L1-L3: lenses, S: Raman sample, F: spectral filter, B: fiber beam-splitter, APD: avalanche photodiode, TDC: time-to-digital converter, D: dichroic. (c) Stokes photon purity measurements and theory for the two collection geometries at varying pump bandwidth.}
	\label{fig:g2exp}
\end{figure*}

\begin{figure}[b]
	\centering
	\includegraphics[width=0.9\linewidth]{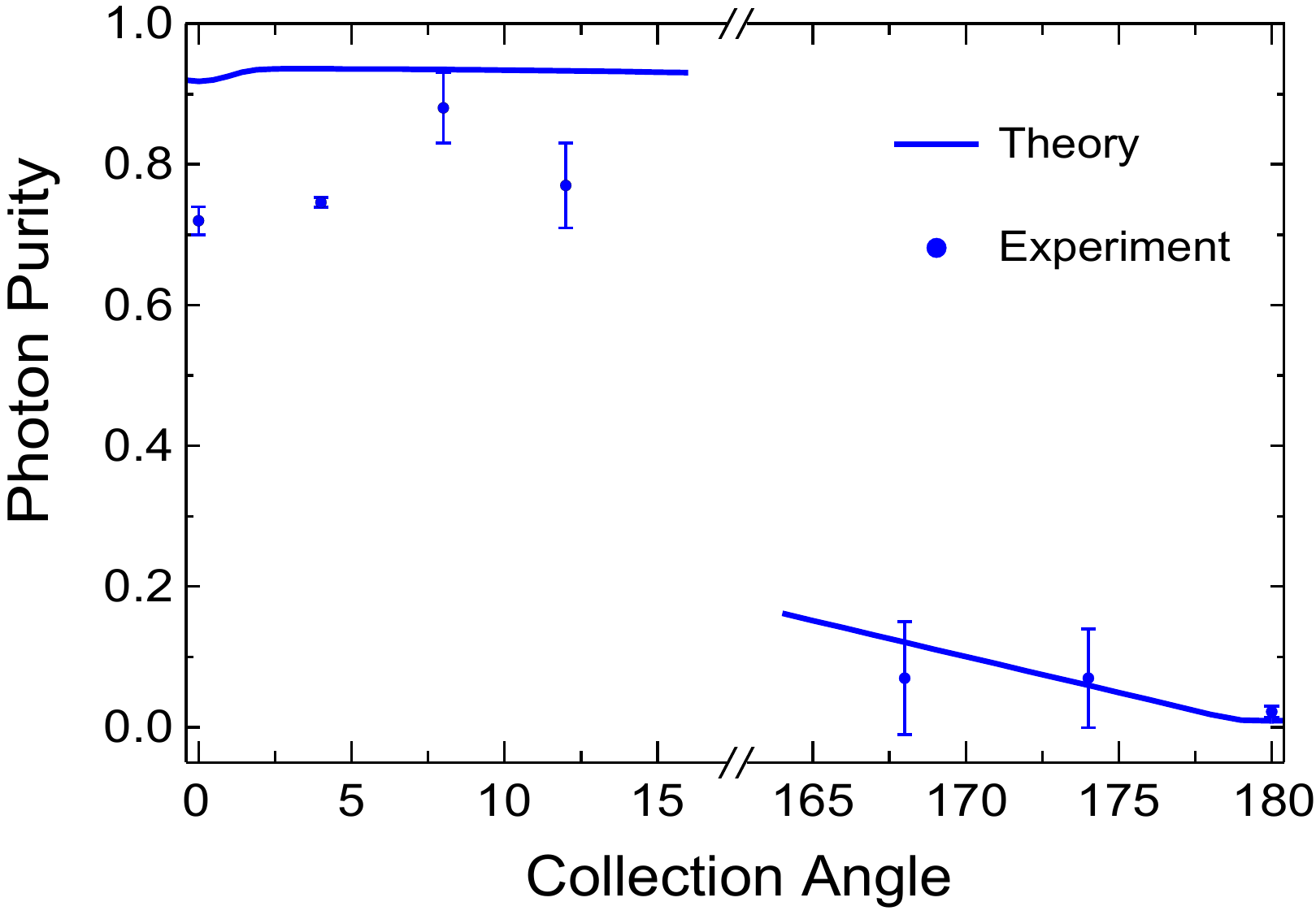}
	\caption{Stokes photon purity measurements for varying collection angle at pump FWHM = 7 nm, along with predictions of the three-dimensional off-axis theory (see text) for pump and collection mode Fresnel numbers $\mathscr{F}=0.1$ corresponding to the experimental parameters.}
	\label{fig:g2expoffaxis}
\end{figure}

In its simplest form the interaction in Eq.~(\ref{eq:Vt}) describes a two-mode squeezing operation of the Stokes and Q-fields, which leads to thermal photon-number statistics of the Stokes field created spontaneously through the Raman interaction~\cite{Mauerer09}. It was found previously that correlations due to chromatic dispersion in the Raman medium and correlations due to the finite lifetime of the Q-field excitations both independently lead to a multimode-nature of this squeezing~\cite{Wasilewski_2006, Raymer_1985}, affecting the photon-number statistics which become more Poissonian as the number of squeezed modes increases \cite{Mauerer09,Christ11}. As we have derived, collection geometry and pump focusing also affect the entanglement of photon and CE and thence the photon purity, multimode nature of the squeezing, and the photon statistics. 

Stokes photon second-order coherence $g^{(2)}$ measurements are performed with the Hanbury-Brown--Twiss interferometers shown in Figs.~\ref{fig:g2exp}(a) and (b), for which $g^{(2)} = N_{12}R/N_1 N_2$, where $N_{12}$ represents coincident detection of two Stokes photons in both arms of the interferometer, $R$ the number of pump pulses over which the counts are taken, and $N_1$ ($N_2$) the counts in arm 1 (2). In the multimode squeezing process as described, the $g^{(2)}$ autocorrelation function takes the form $g^{(2)} = 1 + 1/K$, where $K$ is the effective number of squeezed modes and is related to the purity of the photons by $P= 1/K$ \cite{Mauerer09, Christ11}. 

In our experiments, pump pulses of duration 100 fs from a mode-locked Ti:Sapphire laser at 80~MHz repetition rate pass through a $4f$ spectral pulse shaper before they are focused by a lens of focal length 5~cm, generating spontaneous Stokes photons from a room-temperature, single-crystal, bulk sapphire medium (Ted Pella, Inc.) of length 8~mm centered on the beam waist. The Stokes photons are collected by another (the same) 5~cm focal length lens in the co-(counter-)propagating configuration, and are coupled into a single-mode fiber. The beam waist of our pumping light is 9.5~$\mu$m, and the waist of our collection mode is 9~$\mu$m, corresponding to Fresnel numbers of the pump and collection modes $\mathscr{F}$ ($= \mathscr{F}_p, \mathscr{F}_f$) = 0.1. The scattered photons are registered by Excelitas  SPCM-AQ4C avalanche photodiodes and an IDQuantique time-to-digital converter. Figure \ref{fig:g2exp}(c) shows the results of Stokes photon purity measurements $P=g^{(2)}-1$ and our theoretical predictions for two collection schemes: co-propagating [with setup depicted in Fig.~\ref{fig:g2exp}(a)] and counter-propagating [Fig.~\ref{fig:g2exp}(b)] Stokes and pump pulses. We attribute the source of the discrepancy between our theoretical predictions and experimental data to collection of fluorescent photons in the measurement process, which arise from defects in the crystal lattice and whose emission mode overlaps with the Stokes mode. 
Stokes photons collected counter-propagating from the pump have almost zero purity, indicating strong correlations and spatial entanglement with their corresponding excitations, in agreement with the predictions of our model. 

Figure~\ref{fig:g2expoffaxis} shows the results of our photon purity measurements as a function of collection angle for 7 nm pump spectral FWHM. Again the deviation from theoretical prediction is attributed to background fluorescent photons, which are also emitted over solid angle 4$\pi$. Error bars in Fig.~\ref{fig:g2expoffaxis} and in Fig.~\ref{fig:g2exp}(c) are calculated assuming Poissonian photon counting statistics.

\section{Conclusions}

We have developed a Hamiltonian formalism to describe the interaction between a pump laser pulse, Stokes field and collective excitations in Raman media. 
Using a one-dimensional model we have derived the general form of joint photon--CE states created in the low-gain regime of spontaneous Raman scattering. We have found that the CE linewidth creates energy entanglement between the daughter bosons of the pair and through a separate physical mechanism group delay between pump and Stokes pulses (due to chromatic dispersion in the medium) creates momentum/spatial entanglement; together these two effects lead in general to a maximal photon state purity at finite pumping bandwidth. This one-dimensional model is expanded to include the collection of Stokes photons counter-propagating with the pump, in which case we find timing information is available that has a much more substantial effect than does chromatic dispersion, and results in stronger photon--CE spatial entanglement and degradation of the photon purity. We have extended our theory to include photon--CE pair creation in three dimensions with arbitrary Fresnel numbers of the pump and collected Stokes beams, where we find the Stokes photon quantum state differs from the one-dimensional case only by an effective $z$-dependent correction that serves to apodize the interaction length. Finally, we have derived the correlations between photon and CE in the case of off-axis collection of Stokes photons, revealing non-trivial dependence of the photon purity on both collection angle and the focusing of the pump beam. The theory we have developed in this paper has important consequences for Raman-based quantum protocols that rely on spontaneous scattering and two-photon interference \cite{DLCZ,HOM,Kok07}.

We compare the predictions of our model with experiment and confirm the presence of non-negligible correlations between photon and CE due to a finite excitation linewidth, and strong dependence of the scattered photon purity on collection angle, where photons collected counter-propagating with the pump are found in nearly 
completely mixed states. 

\section*{Acknowledgments}
This work is supported in part by NSF Grant Nos.~1521110, 1640968 and 1806572 and NSF award DMR-1747426. The authors thank Michael Raymer for useful comments, Lance Cooper and Astha Sethi for measuring the Raman spectrum of the sapphire crystal used in the experiment, and Yujie Zhang for helpful discussion.

\section*{Appendix: Photon-Excitation Pair Generation in Three Dimensions}

In our three-dimensional treatment, we consider a focused classical pump beam with Gaussian paraxial field given by a collection of plane-waves with transverse wavevector $\mathbf{q}_p = (q_p^x,q_p^y)$ as
\begin{equation*}\tag{A.1} \begin{split}
{E}_p(\boldsymbol{\rho},z,t) &= \frac{4\pi}{w_p^2}\int d\omega_p d^2\mathbf{q}_p \,  \bigg[\mathcal{E}(\omega_p) e^{-w_p^2\vert\mathbf{q}_p\vert^2/4}e^{i\mathbf{q}_p\cdot\boldsymbol{\rho}}\\  &\hspace{4em}e^{i\left[k(\omega_p) - \vert\mathbf{q}_p
\vert^2/2k(\omega_p)\right]z} e^{-i\omega_p t}\bigg] + \text{h.c.},
\end{split}\end{equation*}

\noindent for Gaussian beam waist $w_p$.

\begin{figure}[H]
	\centering
	\includegraphics[width=\linewidth]{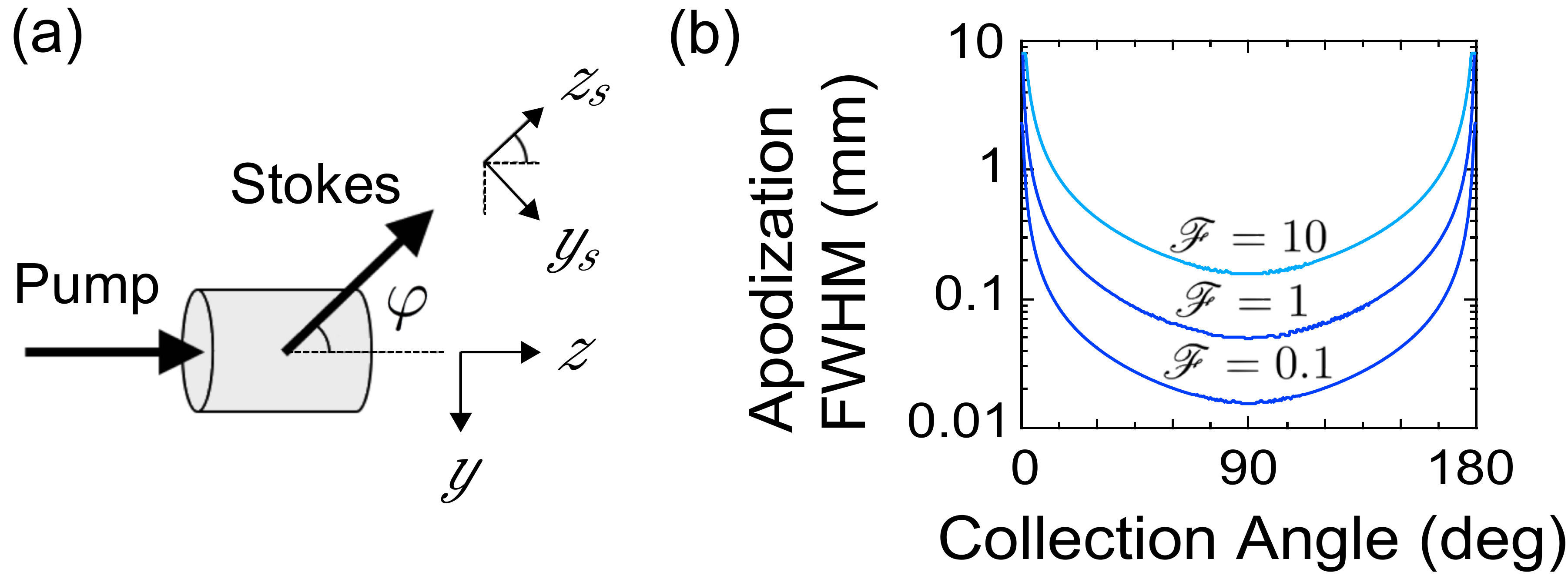}
	\caption{(a) Coordinate system for off-axis collection of Stokes photons. (b) Interaction length apodization function full width at half maximum (FWHM) for varying collection angle, evaluated at pump and collection mode Fresnel numbers $\mathscr{F}=0.1, 1, 10$ for Gaussian beams described in the text.}
	\label{fig:appendix}
\end{figure}

While the Stokes field propagates in all directions, we consider the physical case of collection of photons emitted only around a small range of angles about the axis  $\hat{z}_s$, where we use the following coordinate transformation relative to the $\hat{z}$ axis defined by the pump: 
\stepcounter{equation}
\begin{align}
    \label{eq:xs}\tag{A.2} x_{s} &= x\\
	\tag{A.3} y_{s} &= y \cos\varphi + z\sin\varphi\\
	\label{eq:zs}\tag{A.4} z_{s} &= z \cos\varphi - y\sin\varphi,
\end{align}

\noindent shown schematically in Fig.~\ref{fig:appendix}(a), where $\boldsymbol{\rho_s} = (x_s,y_s)$, and the transverse photon wavevector in the off-axis coordinate system is $\mathbf{q}_s = (q_s^{x},q_s^{y})$. In order to develop the three-dimensional theory, instead of the Stokes photon creation operator considered in the text here we consider the negative frequency component of the paraxial Stokes field operator, defined as
\begin{equation*}\tag{A.5}\begin{split}
\hat{E}_s^{(-)}(\boldsymbol{\rho_s},z_s) &=\\ &\hspace{-5em} -i\sqrt{\hbar\omega_s/2 \mathcal{V} \varepsilon_0} \int d\omega_s d^2\mathbf{q}_s \,  \hat{a}_{\mathbf{q}_s}^\dagger(\omega_s) e^{-i\mathbf{q}_s\cdot\boldsymbol{\rho}_s}  \\ 
&\hspace{4em} \times e^{ - i\left[k(\omega_s) - \vert\mathbf{q}_s\vert^2/2k(\omega_s)\right]z_s},
\end{split}\end{equation*}
\noindent for quantization volume $\mathcal{V}$ and vacuum permittivity $\varepsilon_0$. Then using the transformation in Eqs. (\ref{eq:xs})-(\ref{eq:zs}), we rewrite the Stokes field operator in the original basis as $\hat{E}_s^{(-)}(\boldsymbol{\rho},z) $.
%

We write the interaction term in Eq.~(\ref{eq:Vt}), including the transverse degrees of freedom, as:
\begin{multline*}\label{eq:H_I_3d}
    \tag{A.7}\hspace{-1em}\hat{V}_\textrm{3D}(t) = \gamma_\textrm{3D}\int d^2\boldsymbol{\rho} \int_{-L/2}^{L/2} dz \,  {E}_p(\boldsymbol{\rho},z,t) \hat{E}_s^{(-)}(\boldsymbol{\rho},z)  \\ \hspace{11em} \times \hat{Q}^\dagger(\boldsymbol{\rho},z) + \textrm{h.c.},
\end{multline*}
with the coupling constant $\gamma_\textrm{3D}$ associated with the amplitude of the interaction locally, and $\hat{Q}^\dagger(\boldsymbol{\rho},z)$ given by Eq.~(\ref{eq:Q_rho}) with transverse wavevector $\mathbf{q}_\text{CE} = (q_\textrm{CE}^x,q_\textrm{CE}^y)$. We apply this interaction perturbatively to the vacuum state to find the photon--CE joint state in the paraxial approximation
\begin{multline*}
    \tag{A.8}\hspace{-1em}\ket{\Psi}_\textrm{par} = \\ \mathcal{N}_\textrm{par}\int_{-L/2}^{L/2}dz \int d\omega_s d^2\mathbf{q}_s d\Omega d^2\mathbf{q}_{\mathrm{CE}} \, f_\textrm{par}(\omega_s, \mathbf{q}_s, \Omega, \mathbf{q}_{\mathrm{CE}}, z)\\ \times \hat{a}_{\mathbf{q}_s}^\dagger(\omega_s) \hat{B}^\dagger(\Omega,\mathbf{q}_{\mathrm{CE}},z)\vac,
\end{multline*}

\noindent where we have assumed that the transverse extent of the Raman medium is much larger than the transverse extent of the focused pump beam, thus recovering the transverse momentum conserving relations $q_p^x = q_s^x + q_\text{CE}^x$ and $q_p^y = q_s^y\cos\varphi - [k(\omega_s)-\abs{\mathbf{q}_s}^2/2k(\omega_s)]\sin\varphi + q_\textrm{CE}^y$. 
Here $\mathcal{N}_\textrm{par}$ is a normalization factor
. Keeping terms of $\mathcal{O}\left(\vert\mathbf{q}_s\vert^2/k(\omega_s)^2\right)$ consistent with the paraxial approximation, the JA is given by
\begin{equation*}
    \tag{A.9}f_\textrm{par}(\omega_s, \mathbf{q}_s, \Omega, \mathbf{q}_{\mathrm{CE}}, z) =  \mu(\omega_s, \mathbf{q}_s, \Omega, \mathbf{q}_{\mathrm{CE}}, z) f_\text{1D}(\omega_s,\Omega,z)
\end{equation*}

\vspace{1em}
\noindent where


\begin{align}
    \notag &\mu(\omega_s, \mathbf{q}_s, \Omega, \mathbf{q}_{\mathrm{CE}}, z) =  \\  
    &\notag \ \exp\left[\frac{-w_p^2}{4}\left(q_s^{x} + q^x_\text{CE}\right)^2\right]  \exp\left[\frac{-w_p^2}{4} (q_s^y\cos\varphi + q_\text{CE}^y)^2 \right] \\
    &\notag \ \times\exp\left[\frac{-w_p^2}{4} (k(\omega_s)^2 - \vert\mathbf{q}_s\vert^2)\sin^2\varphi\right]\\
    &\notag \ \times \exp\left[\frac{w_p^2}{2}k(\omega_s)(q_s^y\cos\varphi + q_\text{CE}^y)\sin\varphi \right]\\
    &\notag \ \times \exp\left[-i\frac{(q_s^{x} + q^x_\text{CE})^2}{2k(\omega_s+\Omega)}z\right] \exp\left[-i\frac{(q_s^y\cos\varphi + q_\text{CE}^y)^2}{2k(\omega_s+\Omega)}z\right] \\ 
    &\notag \ \times \exp\left[-i\frac{(k(\omega_s)^2 - \vert\mathbf{q}_s\vert^2)\sin^2\varphi}{2k(\omega_s+\Omega)}z\right]  \\
    &\notag \ \times \exp\left[i\frac{k(\omega_s)(q_s^y\cos\varphi + q_\text{CE}^y)\sin\varphi}{k(\omega_s+\Omega)}z \right] \\
    &\notag \ \times \exp\left[i\left(\frac{\vert\mathbf{q}_s\vert^2\cos\varphi}{2k(\omega_s)} - q_s^{y} \sin\varphi \right)z\right]\\
    \tag{A.9}& \ \times \exp\left[i(k(\omega_s) - k(\omega_s)\cos\varphi) z\right]
\end{align}

\noindent and $f_\textrm{1D}(\omega_s,\Omega,z)$ is given by Eq.~(\ref{eq:f1D}) for the one-dimensional case. Considering the physical case of Stokes 

\noindent photons collected by a lens and coupled to a single-mode fiber, assuming that the projection of the spatial mode supported by the fiber onto free space by the lens can be well approximated by a Gaussian, this configuration projects the scattered Stokes photons onto a state with creation operator $\hat{A}_s^{\dag}(\omega_s)$ given in Eq.~(\ref{eq:A_s}). Letting $\ket{u_f(\omega_{s})} = \hat{A}_s^{\dag}(\omega_s)\vac$, the projection of the emitted state onto this concentric collection mode, given by normalizing the state $\int d\omega_s\, \ket{u_f(\omega_{s})}\braket{u_f(\omega_{s})}{\Psi}_\textrm{par}$, results in a state with the joint amplitudes given by Eqs.~(\ref{eq:f_3D}) and (\ref{eq:beta_phi}). 

In the three-dimensional case this collection scheme leads to the apodization function in the reduced density matrix of the Stokes photon $\alpha(z,\varphi)$ given in Eq.~(\ref{3Doffaxis_alpha}). The behavior of the apodization function FWHM for varying Fresnel number is shown in Fig.~\ref{fig:appendix}(b) for the same bulk Al$_2$O$_3$ medium considered in the text. For a fixed medium length and collection angle, a more tightly focused pump beam will generate a narrower apodization function, resulting in the increase of the collected photon purity.

\bibliographystyle{apsrev4-2}
\bibliography{PhotonMatter_manuscript.bib}

\end{document}